\documentclass[final, 5p, times, twocolumn]{elsarticle}

\usepackage[colorlinks=true,linkcolor=blue]{hyperref}%
\usepackage{amsmath}
\usepackage{graphicx}
\usepackage[utf8]{inputenc}
\usepackage{url}
\usepackage{bm}
\usepackage{booktabs}
\usepackage{lineno}
\usepackage{ulem}
\usepackage{listings}
\usepackage[dvipsnames]{xcolor}
\usepackage{subfig}
\usepackage[capitalise]{cleveref}
\usepackage{tabularx}
\usepackage{lmodern}
\usepackage{gensymb}
\usepackage{fontawesome}
\newcommand{\taurunner}{\texttt{TauRunner}}
\newcommand{\Particle}{\texttt{Particle}}
\newcommand{\Track}{\texttt{Track}}
\newcommand{\Body}{\texttt{Body}}
\newcommand{\XS}{\texttt{CrossSection}}

\definecolor{codegreen}{rgb}{0,0.6,0}
\definecolor{codegray}{rgb}{0.5,0.5,0.5}
\definecolor{codepurple}{rgb}{0.58,0,0.82}
\definecolor{backcolour}{rgb}{0.95,0.95,0.92}

\lstdefinestyle{mystyle}{
    backgroundcolor=\color{backcolour},   
    commentstyle=\color{codegray},
    keywordstyle=\color{codegreen},
    numberstyle=\tiny\color{codegray},
    stringstyle=\color{orange},
    basicstyle=\ttfamily\footnotesize,
    breakatwhitespace=false,         
    breaklines=true,                 
    captionpos=b,                    
    keepspaces=true,                 
    numbers=left,                    
    numbersep=5pt,                  
    showspaces=false,                
    showstringspaces=false,
    showtabs=false,                  
    tabsize=2
}

\journal{Computer Physics Communications}
\begin{document}
\begin{frontmatter}
\title{\texttt{TauRunner}: A Public Python Program to Propagate Neutral and Charged Leptons {\raisebox{-2.50\depth}{\centering{\href{https://github.com/icecube/TauRunner}{\huge\color{BlueViolet}\faGithub}}}}}

\author[a,b]{Ibrahim~Safa\corref{correspondingauthor}}
\author[a,b]{Jeffrey~Lazar\corref{correspondingauthor}}
\author[b]{Alex~Pizzuto}
\author[a]{Oswaldo~Vasquez}
\author[a,c]{\\Carlos~A.~Arg{\"u}elles} 
\author[b]{Justin~Vandenbroucke}

\cortext[correspondingauthor]{Corresponding authors: \url{isafa@fas.harvard.edu},\\\url{jeffreylazar@fas.harvard.edu}}

\address[a]{Department of Physics \& Laboratory for Particle Physics and Cosmology, Harvard University, Cambridge, MA 02138, USA}
\address[b]{Department of Physics and Wisconsin IceCube Particle Astrophysics Center, University of Wisconsin-Madison, Madison, WI 53706, USA}
\address[c]{The NSF AI Institute for Artificial Intelligence and Fundamental Interactions}

\begin{abstract}
In the past decade IceCube's observations have revealed a flux of astrophysical neutrinos extending to $10^{7}~\rm{GeV}$.
The forthcoming generation of neutrino observatories promises to grant further insight into the high-energy neutrino sky, with sensitivity reaching energies up to $10^{12}~\rm{GeV}$.
At such high energies, a new set of effects becomes relevant, which was not accounted for in the last generation of neutrino propagation software.
Thus, it is important to develop new simulations which efficiently and accurately model lepton behavior at this scale.
We present \taurunner{}, a \texttt{PYTHON}-based package that propagates neutral and charged leptons.
\taurunner{} supports propagation between $10~\rm{GeV}$ and $10^{12}~\rm{GeV}$.
The package accounts for all relevant secondary neutrinos produced in charged-current tau neutrino interactions.
Additionally, tau energy losses of taus produced in neutrino interactions is taken into account, and treated stochastically.
Finally, \taurunner{} is broadly adaptable to divers experimental setups, allowing for user-specified trajectories and propagation media, neutrino cross sections, and initial spectra.
\end{abstract}

\end{frontmatter}

\section{Introduction}
\label{sec:intro}
Most natural and anthropogenic neutrino sources produce neutrinos with energies below 1~TeV~\cite{Vitagliano:2019yzm}, where the smallness of the neutrino-nucleon cross section~\cite{Formaggio:2012cpf} allows them to freely stream through large amounts of column density.
Famously, low-energy solar neutrinos produced in nuclear processes in the Sun are not only able to escape the dense solar core but also can diametrically traverse hundreds of Earths unimpeded. 
In this energy range, the negligible scattering rates imply that the problem of neutrino transport requires only considering the changing of flavors between neutrinos.
This problem prompted the neutrino community to develop analytical methods and numerical schemes to compute the neutrino oscillation probabilities efficiently~\cite{Barenboim:2019pfp}, \textit{e.g.} \texttt{nuSQuIDS}~\cite{Arguelles:2020hss} among others~\cite{Huber:2007ji,Calland:2013vaa,Wallraff:2014vl,Arguelles:2019phs}.
These solutions, currently available through a variety of software packages and libraries~\cite{prob3pp,nusquids}, are currently used by neutrino experiments to extract the neutrino oscillation parameters.

Recently, the construction of gigaton-scale neutrino detectors, such as the IceCube Neutrino Observatory~\cite{IceCube:2016zyt} in the Antarctic continent, has enabled the observation of neutrinos with energies as large as 10~PeV.
In this high-energy frontier, neutrino oscillations can be safely neglected for Earth-traversing neutrinos; however, in this regime, the neutrino interaction length becomes comparable to or much smaller than Earth's diameter~\cite{Gandhi:1998ri}, requiring new solutions to the neutrino transport problem.
While the first generation of software packages that aimed to address this problem~\cite{Gazizov:2004va,DeYoung:865626,Yoshida:2003js,Arguelles:2020hss,Vincent:2017svp,shigeruyoshida_2020_4018117} included the effects of neutrino-nucleon neutral- and charged-current interactions, they neglected secondary neutrinos from lepton charged-current interactions, except in the case of tau neutrinos.
Tau neutrinos were handled as a special case because, as recognized in~\cite{Halzen:1998be}, due to the short lifetime of the taus, it still carries most of its energy at the time of decay, yielding high-energy secondary neutrinos.
This effect, often known as tau regeneration, implies that Earth is less opaque to tau neutrinos relative to other flavors.

In these first-generation packages tau regeneration was implemented by using the so-called on-spot tau decay approximation, which neglects tau energy losses.
Though this approximation satisfies the needs of most current scenarios and experimental settings, next-generation neutrino telescopes aim to reach EeV energies~\cite{IceCube-Gen2:2020qha,POEMMA:2020ykm}.
At these extremely high energies, the taus produced in neutrino interactions are sufficiently long-lived that their energy losses cannot be neglected.
Recently, dedicated software packages have been made available to solve this problem in this energy regime. 
However, the bulk of the available solutions neglects the stochasticity of tau losses considering only their mean effect.
This limits their ability to function as event generators in neutrino telescopes and produces mismodeling of the yield of tau-induced events for a small number of scatterings, where the stochastic nature of the losses is more relevant.
A notable exception is the \texttt{NuPropEarth}~\cite{Garcia:2020jwr} package developed for the KM3NeT experiment~\cite{Adrian-Martinez:2016fdl}, which is presently being built in the Mediterranean Sea.
Though \texttt{NuPropEarth} offers a complete solution, this package requires a large number of dependencies to function, making its distribution and installation difficult.

In this article, we describe a new package, \taurunner{}, that aims to provide a complete and versatile solution to the neutrino transport problem at high energies.
Our python-based package is designed to have minimal dependencies, allow the user to construct arbitrary neutrino trajectories and propagation media, and provide interfaces to modify physics inputs such as neutrino cross sections easily.
This package was first introduced in~\cite{Safa:2019ege,Vazquez:2021txv}, where it was used to study the ANITA anomalous events~\cite{Gorham:2016zah, Gorham:2018ydl}, and is currently used in studies relating to extremely high-energy neutrinos in IceCube~\cite{IceCube:2021pue}.
With respect to the preliminary version, the version presented in this paper contains significant improvements in terms of performance and available features to the user. 
In this article, we describe the software and provide examples, benchmarks and comparisons to other packages that have similar aims.
We expect that our software will be useful for next-generation neutrino detectors operating in liquid water (P-ONE~\cite{P-ONE:2020ljt}), solid water (IceCube-Gen2~\cite{IceCube-Gen2:2020qha}), mountains (Ashra NTA~\cite{Sasaki:2017zwd}, TAMBO~\cite{Romero-Wolf:2020pzh}), and outer space (POEMMA~\cite{POEMMA:2020ykm}).
Our hope is that the success of neutrino oscillation measurements enabled by the previous generation of software will be mirrored in the study of high-energy neutrino properties with efficient propagation software such as the one presented in this paper.

The rest of this article is organized as follows.
In Sec.~\ref{sec:algorithm} we outline the transport equation, the algorithm used to solve it, and the interaction; in Sec.~\ref{sec:code} we explain the code structure; in Sec.~\ref{sec:performance} we present studies of the software performance; in Sec.~\ref{sec:examples} we lay out the examples included with the code.
Finally in Sec.~\ref{sec:conclusion} we conclude.

\renewcommand{\arraystretch}{1.5}
\begin{table*}[!ht]
    \centering
    \begin{tabularx}{\textwidth}{ l | c c c c c }
        \hline
        \hline
        \textbf{Software} & \textbf{Language} & \textbf{Input} & \textbf{Output} & \textbf{Medium} & \textbf{Energy losses$\left(l^{\pm}\right)$} \\
        \hline
        \taurunner{} & \texttt{Python} & $\nu_{\tau,\mu,e}, \tau, \mu$ & $\nu_{\tau,\mu,e}, \tau, \mu$ & Earth/Sun/Moon/Custom & \texttt{PROPOSAL} \\
        \hline
        \texttt{NuPropEarth}\cite{Garcia:2020jwr}  & \texttt{C$++$} & $\nu_{\tau,\mu,e} $ & $\nu_{\tau,\mu,e}, \tau$ & Earth/Custom & \texttt{TAUSIC} \\ 
        \hline
        \texttt{nuPyProp}\cite{NuSpaceSim:2021hgs} & \texttt{Python}/\texttt{FORTRAN} & $\nu_{\tau}$ & $\tau$ & Earth & Internal \\ 
        \hline
        \texttt{NuTauSim}\cite{Alvarez-Muniz:2018owm} & \texttt{C$++$} & $\nu_{\tau}$ & $\tau$ & Earth & Continuous   \\ 
        \hline
        \hline
    \end{tabularx}
    \caption{\textbf{\textit{Software comparison table.}} Each row of this table represents a given package. Input and output particles include their not explicitly mentioned antiparticles. 
    Custom medium refers to a user-defined \Body{} in \taurunner{}. The Energy losses column compares the treatment of charged particle losses.
    }
    \label{tab:softable}
\end{table*}

\section{Algorithm Overview}
\label{sec:algorithm}
The aim of this software is to solve the transport equation for high-energy neutrino fluxes passing through matter.
The transport equation can be written as follows~\cite{GonzalezGarcia:2005xw},
\begin{equation}
    \label{eq:transport}
    \frac{d \vec{\varphi}(E, x)}{d x}=-\sigma(E) \vec{\varphi}(E, x)+\int_{E}^{\infty} d \tilde{E} ~  f(\tilde{E}, E) \vec{\varphi}(\tilde{E}, x),
\end{equation}
where $E$ is the neutrino energy, $x$ is the target column density, $\sigma(E) = {\rm{diag}}(\sigma_\nu, \sigma_{\bar\nu})$ holds the total  $\nu$ and $\bar{\nu}$ cross section per target nucleon, $f(\tilde{E}, E)$ is a function that encodes the migration from higher to lower neutrino energies and between $\nu$ and $\bar{\nu}$, and $\vec{\varphi}(E, x) = \{\phi_\nu, \phi_{\bar\nu}\}$ contains the neutrino and anti-neutrino spectrum.
At energies supported by this package, $10$ GeV--$10^{12}$ GeV, neutrino-nucleon deep inelastic scattering (DIS) is the dominant neutrino interaction process.
The first term on the right hand side accounts for the loss of flux at energy $E$ due to charged-current (CC) and neutral-current (NC) interactions, whereas the second term is the added contribution from neutrinos at higher energy, $\tilde{E}$, to $E$ through NC interactions of $\nu_{e, \mu, \tau}$ and CC interactions in the $\nu_{\tau}$ channel.

This latter channel is unique in that the short $\tau$ lifetime causes the decay of the charged lepton before losing a large fraction of the parent energy.
The $\tau$ then decays into a daughter $\nu_{\tau}$, meaning that the primary $\nu_{\tau}$ flux is not lost, but only cascades down in energy.
Moreover, if the $\tau$ decays leptonically, $\bar{\nu}_{\mu}$ and $\bar{\nu}_{e}$ are created, contributing significantly to the outgoing flux. 
By default, \taurunner{} takes all those contributions into account.
The story is simpler for the electron channel.
There, CC interactions result in electrons which lose their energy quickly and are subsequently absorbed in the medium.
As a result, electron losses are not modeled in \taurunner{} by default, though the capability exists if needed.
For the muon flavor, muons resulting from CC interactions can travel $\mathcal{O}(1)$~kmwe. 
Therefore, it is important to model the propagation and losses of muons near the point of exit, and that is accounted for in \taurunner{} as well.

\subsection{Algorithm Description}

\begin{figure}[t]
    \centering
    \includegraphics[width=0.48\textwidth]{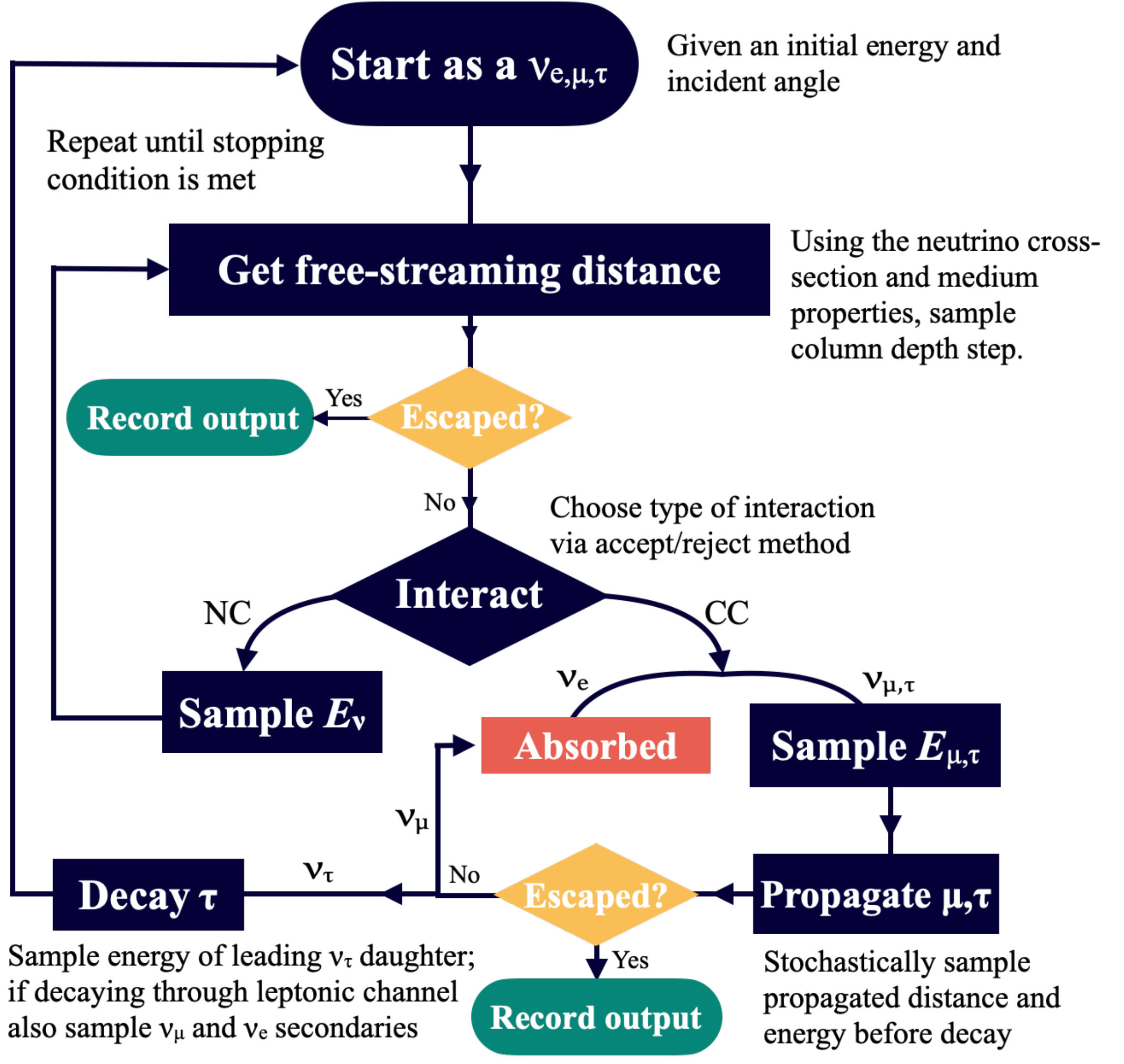}
    \caption{\textit{\textbf{Flowchart of the \taurunner{} propagation algorithm.}}
    Square boxes indicate actions performed by the software. 
    Diamond boxes indicate decision-making stopping points.
    Rounded-corner squared boxes indicate beginning and end of the algorithm.
    Note that users can select also charged leptons as the initial state, in which case
    the algorithm skips straight to the charged particle propagation step.
    }
    \label{fig:flowchart}
\end{figure}

In \taurunner{}, Eq.~\eqref{eq:transport} is solved using a Monte-Carlo approach.
A flowchart of the \taurunner{} Monte-Carlo algorithm is shown in Fig.~\ref{fig:flowchart}. 
Given an initial neutrino type, energy, and incident angle, it begins by calculating the mean interaction column depth, $\lambda_{\rm{int}}$, which depends on the medium properties and neutrino cross section.
A column depth is then randomly sampled from an exponential distribution with parameter $\lambda_{\rm{int}}$, and the neutrino advances the corresponding free-streaming distance. 
If the neutrino does not escape the medium, either an NC or CC interaction is chosen via the accept/reject method.
In the case of an NC interaction, the neutrino energy loss is sampled from the differential cross section, and the process repeats.
In the case of a CC interaction, a charged lepton is created with energy sampled from the neutrino differential cross section.

The treatment of the charged lepton then varies according to the initial neutrino flavor.
Electrons are assumed to be absorbed and the propagation stops there. 
$\mu$ and $\tau$, however, are recorded and passed to \texttt{PROPOSAL} to be propagated through the same medium.
$\mu$ that do not escape will either decay at rest resulting in neutrinos that are below the energies supported by \taurunner{}, or get absorbed.
Therefore a $\mu$ that does not escape is not tracked further. 
Finally, $\tau$s can either escape or decay.
In the latter case, a secondary $\nu_{\tau}$ is created whose energy is sampled from tau decay distributions provided in~\cite{Dutta:2002zc}.
Additionally, if the $\tau$ decays leptonically, $\nu_{e}$ or $\nu_{\mu}$ will be created.
When this happens, the properties of the resulting secondaries are recorded and added to a basket which stores all secondary particles to be propagated together after the primary particle propagation is complete.

\subsection{Lepton Interactions and Decays}

Measurements of neutrino cross sections with matter have been performed up to a few PeV in energy~\cite{Zyla:2020zbs}.
This includes a multitude of accelerator~\cite{AguilarArevalo:2010zc,Tzanov:2005kr} and reactor~\cite{Vogel:1999zy,Kurylov:2002vj} experiments as well as solar~\cite{Agostini:2018uly}, atmospheric~\cite{Li:2017dbe}, and astrophysical neutrinos~\cite{Aartsen:2017kpd,IceCube:2020rnc}.
However, the energy range supported by \taurunner{} goes far beyond the measurements, where the fractional momenta, $x_{\rm{Bjorken}}$, of the quarks probed by the neutrino can reach $x_{\rm{Bjorken}}\ll10^{-8}$.
The nucleon structure function is not measured at such low $x_{\rm{Bjorken}}$ and is extrapolated in cross section calculations~\cite{CooperSarkar:2011pa,Garcia:2020jwr}. 
Such extrapolations neglect gluon color screening making perturbative QCD calculations of the neutrino cross section grow faster than allowed by unitarity at extremely high energies~\cite{Froissart:1961ux}.
Phenomenological approaches to include gluon screening parameterize the extremely small $x_{\rm{Bjorken}}$ behavior using a dipole model~\cite{Arguelles:2015wba} of the nucleon so as to result in a $\ln^2(s)$ dependence of the cross section at extremely high energies~\cite{Block:2011vz}.
This ultimately results in a difference of a factor $\sim2$ at $10^{12}$ GeV.
\taurunner{} provides, by default, neutrino and anti-neutrino DIS cross section tables for two PDF models: a perturbative QCD calculation~\cite{CooperSarkar:2011pa}, and a dipole model~\cite{Arguelles:2015wba}.
The user also has the option to provide their own cross sections, see Sec.~\ref{subsec:xs} for more details.

In the Standard Model, when neutrinos undergo CC interactions, they convert to their charged partners through the exchange of a $W$ boson.
Charged particles lose energy in dense media through many processes, and the relative importance of each process depends on the lepton's mass and its energy~\cite{ParticleDataGroup:1994kdp}.
At lower energies, a charged lepton can ionize atoms as it traverses the medium.
This process is described by the Bethe-Bloche equation, and at higher energies scales logarithmically and becomes sub-dominant for all flavors.
A charged lepton can also interact with the electric field of a nucleus, losing energy in the process through the emission of a photon.
This process, called bremsstraahlung, scales like the inverse-sqaured mass of the lepton, and is therefore the dominant energy loss mechanism for electrons.
Another possible interaction with the field of a nucleus leads to the production of electron-positron pairs.
This process scales like the inverse of the lepton mass, and is one of the leading energy-loss mechanisms for $\mu$ and $\tau$.
Finally, the leptons can also lose energy by exchanging a photon with a nucleon, in what is referred to as a photonuclear interaction.
This process dominates tau energy losses at the highest energies ($\geq 10^{9}$~GeV).
The aforementioned processes are implemented in \texttt{PROPOSAL}~\cite{Koehne:2013gpa}, which we use to model them in \taurunner{}.
Apart from interacting, $\mu$ and taus can also undergo weak decays.
This process scales like the mass of the lepton to the fifth power, and is therefore the most likely outcome for taus propagating in Earth up to $10^{9}$~GeV.
Above this energy, the total interaction length for other processes becomes shorter than the decay length.
$\mu$, on the other hand, are much more likely to lose all of their energy before decaying at rest, or getting absorbed by a nucleus.
Therefore, we only model decays of $\tau$ leptons using parametrizations in~\cite{Dutta:2002zc}.

\section{Structure of the Code}
\label{sec:code}
\taurunner{} may be run either from the command line by running \texttt{main.py} or may be imported to run within another script or \texttt{Jupyter} notebook.
To run from the command line, the user must minimally specify the initial energy, the incident nadir angle, and the number of events simulate.
These can be specified with the \texttt{-e}, \texttt{-t}, and \texttt{-n} command line flags respectively.
This will run the \taurunner{} algorithm in Earth with a chord geometry.
The \taurunner{} output will be printed in the terminal unless an output file is specified with the \texttt{--save} flag.
If this option is specified, \taurunner{} will save both a \texttt{numpy} array and a \texttt{json} file with the configuration parameters at the specified location.
In order to ensure reproducibility, the user may specify a seed for the random number generator with the \texttt{-s} flag.
By default, \texttt{main.py} propagates an initial $\nu_{\tau}$ flux, but a user may specify other initial particle types by using the \texttt{--flavor} flag.
Additional options that may be specified by the user can be found in the \texttt{initialize\_args} function of \texttt{main.py} or by running \texttt{main.py} with the \texttt{-h} flag.

To run within another script or \texttt{Jupyter} notebook the user must import the \texttt{run\_MC} function from \texttt{main.py}.
In this latter case one must also create a \taurunner{} \Particle{}, \Track{}, \Body{}, \XS{} objects and a \texttt{PROPOSAL} propagator.
The \Particle{} class, described in Sec.~\ref{subsec:particle}, contains the particle properties as well as methods for particle propagation.
The \Track{} class, described in Sec.~\ref{subsec:track}, parametrizes the geometry of the particle trajectories.
The \Body{} class, described in Sec.~\ref{subsec:body}, defines the medium in which the propagation is to occur.
The \XS{} class, described in Sec.~\ref{subsec:xs}, defines neutrino cross section model.
Additionally, \taurunner{} provides a convenience function for constructing \texttt{PROPOSAL} propagators, \texttt{make\_propagator}, which can be imported from the \texttt{utils} module.
Explicit examples of how to run \taurunner{} can be found in Sec.~\ref{sec:examples}.
\texttt{Casino.py} combines these classes according to the logic outlined in Fig.~\ref{fig:flowchart}.

After discussing the package broadly, we will discuss conventions in Sec.~\ref{subsec:convent} and describe \taurunner{}'s output in Sec.~\ref{subsec:output}

\subsection{\Particle{}}
\label{subsec:particle}

A \Particle{} instance contains the structure of a \taurunner{} event.
This includes, among other quantities, the particle's initial and current energies, particle type, and position.
Additionally, it has a number of methods for particle decay and interaction as well as charged lepton propagation.
Finally, the $\tau$ decay parametrization is contained in \texttt{particle/utils.py}.

The user may propagate $\nu_{e}$, $\nu_{\mu}$, $\nu_{\tau}$, $\mu^{-}$, $\tau^{-}$, or any of the corresponding anti-particles in \taurunner{}.
To do this, the user should initialize the the \Particle{} object with the corresponding Particle Data Group Monte Carlo number~\cite{ParticleDataGroup:1994kdp}.
It should be noted that the user may create an $e^{\pm}$, but the internal logic of \taurunner{} assumes all $e^{\pm}$ are immediately absorbed and thus no propagation occurs; see Fig.~\ref{fig:flowchart}.

\subsection{\Track{}}
\label{subsec:track}

\begin{figure}[t]
    \centering
    \subfloat[]{\includegraphics[width=.35\textwidth]{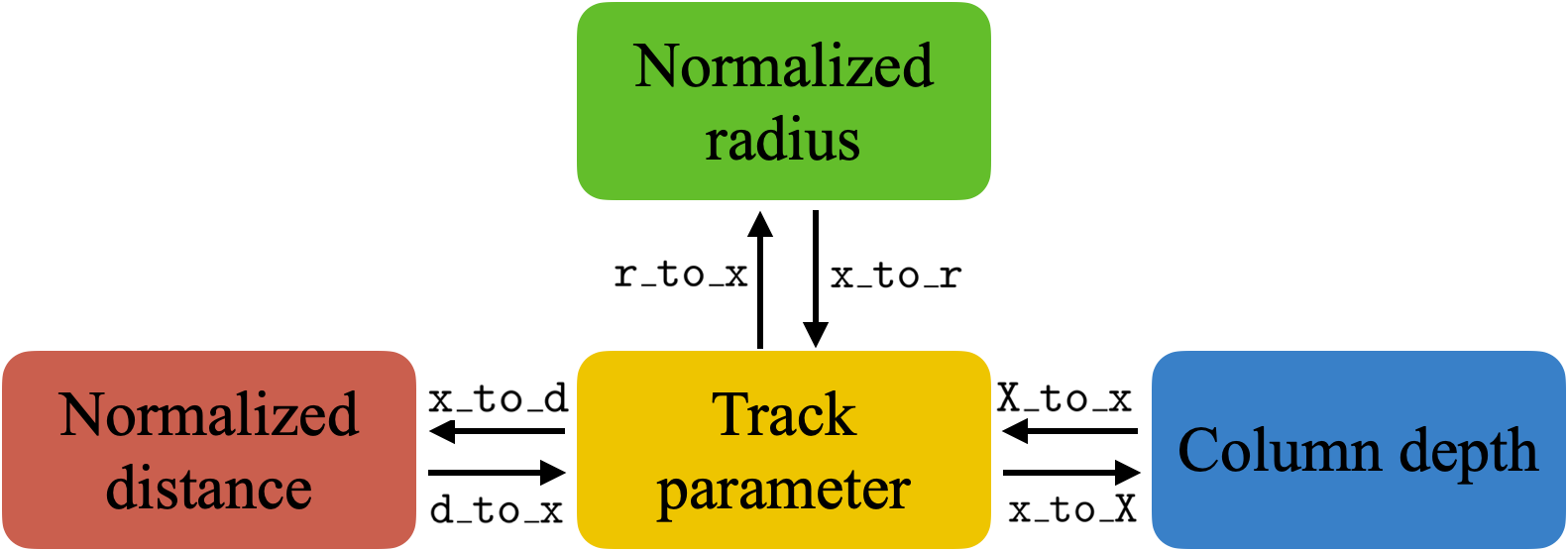}}\\
   \subfloat[]{\includegraphics[width=0.35\textwidth]{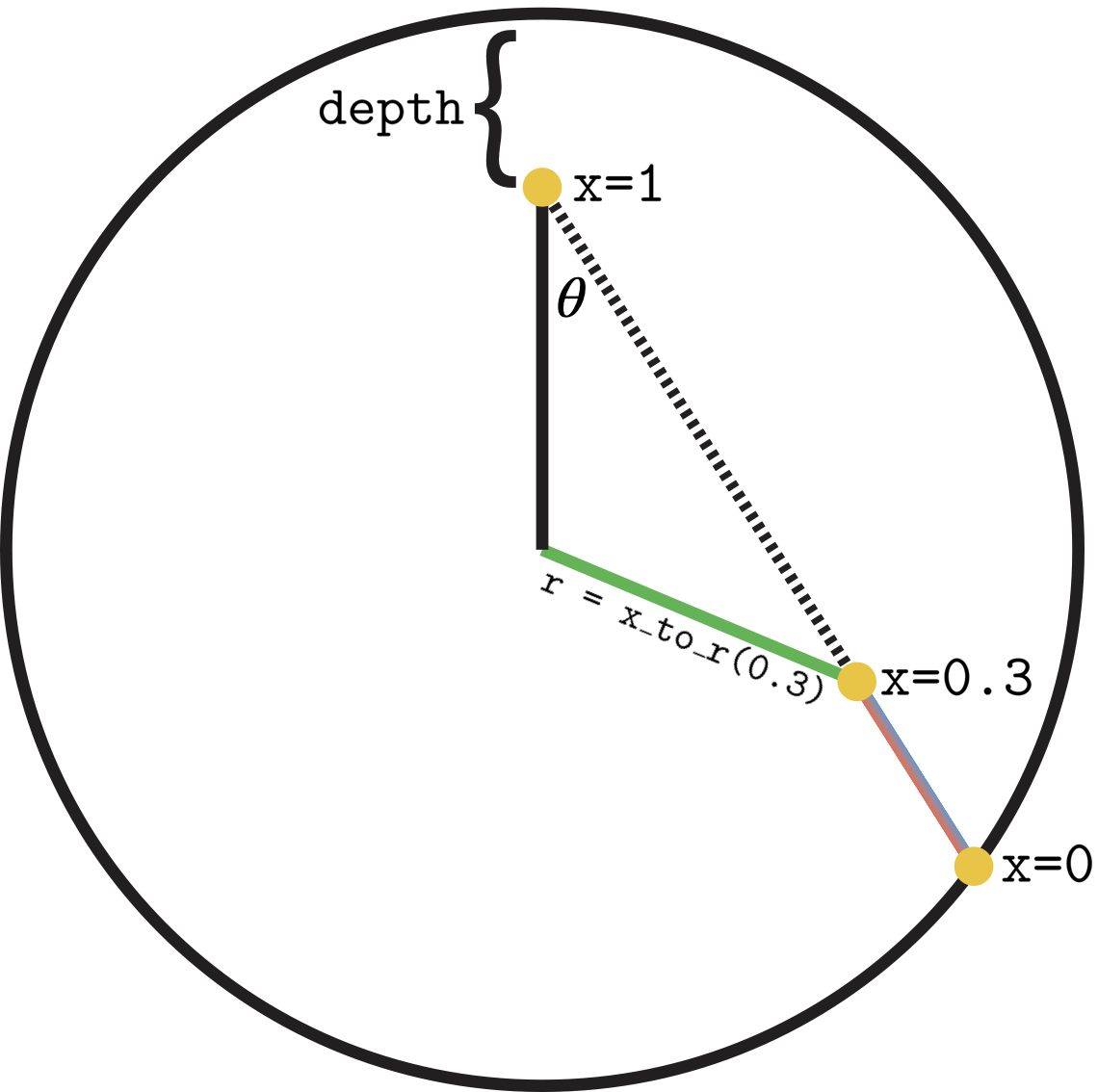}}\\
    \caption{\textbf{\textit{Schematic of \taurunner{} geometry as contained within the \Track{} class.}} (a) shows the relation between the physical quantities relevant to propagation and the affine parameter that parametrizes the \Track{}. The arrows connecting these quantities are labeled with the functions used to convert between them in \taurunner{}. Specifically, these are the functions a user must define in order to specify a custom \Track{} geometry. All distances are normalized with respect to the radius of the body in which the track sits. (b) shows a diagram of these parameters  within a spherical \taurunner{} body. Colors correspond to the boxes in (a). Additionally, it illustrates the \texttt{depth} parameter which intuitively gives the depth of the detector.}
    \label{fig:track_diagram}
\end{figure}

The \Track{} class contains the geometrical information about the particle's trajectory.
A track is parametrized by an affine parameter which defines the position along the trajectory: 0 is the beginning of the trajectory, and 1 is the end.
Almost all of the methods of the \Track{} class are mappings between the affine parameter and physically relevant quantities, \textit{e.g.} radius, distance traveled, and column depth.
The only argument which is generic to the \Track{} class is \texttt{depth} which specifies the distance below the surface of the body at which to stop propagation.
This may intuitively be thought of as the depth of the detector to which the particles are propagated.
An illustration of the \taurunner{} geometry and a diagram of the functional relation of physical quantities to the affine parameter is shown in Fig.~\ref{fig:track_diagram}

The \Track{} class allows the user to make custom trajectories.
The user need only specify mappings between the affine parameter and these variables.
Different trajectories may require additional arguments from the user, depending on the nature of the trajectory.
To illustrate this point, we can look at the two tracks which are implemented by default, the \texttt{Chord} and \texttt{Radial} trajectories.
The former is used for paths which originate outside the \Body{} and cross a section of \Body{}.
The latter is used for paths which originate at the center of the \Body{}.
The former \texttt{Track} describes neutrinos coming from space and passing through Earth on the way to a detector, as in the case of Earth-skimming $\tau$ searches, while the latter gives the trajectory of a neutrino originating in the center of the planet, relevant for searches for neutrinos from gravitationally trapped dark matter.
Clearly, an incoming angle needs to be specified for the \texttt{Chord} trajectory.
Thus, we can see that the necessary arguments for specifying a \Track{} may vary from one geometry to another.

\subsection{\Body{}}
\label{subsec:body}

The \Body{} class specifies the medium in which the \Particle{} is to be propagated.
In \taurunner{}, we require that all bodies be spherically symmetric, and so a \Body{} may be minimally specified by a physical radius, and a density profile.
The density profile may be a positive scalar, a unary function which returns a positive scalar, or a potentially-mixed list of positive scalars and such functions.
The sole argument of the functions used to specify the density should be the radius at which the density is to be given, in units of the radius of the body, \textit{i.e.} the domains should be $[0,1]$.
In this system $r=0$ is the center of the body and $r=1$ the surface.
If the user wishes to make a layered body, \textit{i.e.} one where a list specifies the density profile, they must pass a list of tuple with the length of this list equal to the number of layer.
The first element of each tuple should be the scalar or function which gives the density, and the second element should be the right hand boundary of the layer in units of the radius.
The last right hand boundary should always be 1 since $r=1$ is the outer edge of the body.
Lastly, all densities should be specified in $\rm{g}/\rm{cm}^{3}$.

In addition to a radius and a density profile, the user may also provide the \texttt{proton\_fraction} argument to specify the fraction of protons to total nucleons in the body.
By default, we assume that the propagation medium is isoscalar, \textit{i.e.} we set the proton fraction to 0.5 throughout the entire body.
As in the case of the density profile, this argument may be a scalar, a function, or a list of function-boundary tuples.
The domains of any functions provided must be [0, 1], and the ranges must be in this same interval.

While the user can construct bodies themselves, there are five bodies implemented by default in \taurunner{}: the Earth, a high-metallicity Sun, and low-metallicity Sun, the moon, a constant density slab.
We use the PREM parametrization to model the densities of Earth~\cite{Dziewonski:1981xy}.
For the Sun, we use fits provided by~\cite{gwensun}.
To instantiate the \texttt{Earth} object, one calls the \texttt{construct\_earth} function, which returns an \texttt{Earth} object.
Additionally, this function allows one to pass in a list of additional layers which will be placed radially outward from the edge of the PREM Earth.
This functionality may be useful for \textit{e.g.} adding a layer of water or ice or adding the atmosphere for simulating atmospheric air showers.
Examples on using this functionality may be found in Sec.~\ref{subsec:body}.
To initialize the Sun, one can use the \texttt{construct\_sun} function.
With this function, the user may specify `HZ\_Sun' or `LZ\_Sun' to use the high- and low-metallicity \taurunner{} suns respectively, or a path to a user defined solar model.
An example of how to input solar models is given in Ex.~\ref{app:solar_model}

\subsection{\XS}
\label{subsec:xs}

The \taurunner{} cross sections module defines the neutrino interactions.
Internally, \taurunner{} assumes that cross sections are equal for all neutrino flavors.
Additionally, \taurunner{} uses the isoscalar approximation by default, \textit{i.e.} it assumes a medium is made of equal parts $p^{+}$ and $n$; however, this assumption may be changed by altering the \texttt{proton\_fraction} of the \Body{} object.
See Sec.~\ref{subsec:body} for more information.
The software includes both CSMS~\cite{CooperSarkar:2011pa} and dipole~\cite{Block2014ConnectionOT} cross sections implemented by default; however, it is straightforward for the user to implement other cross section models by providing \texttt{scipy} splines in the appropriate format.
For the total neutrino cross section these splines are \texttt{scipy.interpolate.UnivariateSpline} objects whose $x$-axis is the $\log_{10}$ of the neutrino energy in eV and whose $y$-axis is the $\log_{10}$ of cross section in $\rm{cm}^{2}$.
The differential cross section splines are \texttt{scipy.interpolate.RectBivariateSpline} objects whose $x$-axis is the $\log_{10}$ of the neutrino energy in eV, whose $y$-axis is a convenience variable which combines the incoming and outgoing neutrino energies, $E_{\rm{in}}$ and $E_{\rm{out}}$, given by
$$\eta=\frac{E_{\rm{out}}-10^{9}~{\rm{eV}}}{E_{\rm{in}}-10^{9}~{\rm{eV}}},$$
and whose $z$-axis is the $\log_{10}$ of incoming energy times the differential cross section in $\rm{cm}^{2}$.
An example of how to construct these splines is given in Ex.~\ref{app:xs}.

\subsection{\texttt{PROPOSAL}}
\label{subsec:proposal}

To propagate charged leptons, \taurunner{} relies on \texttt{PROPOSAL}, an open source C++ program with python bindings.
A utility module to interface with \texttt{PROPOSAL}, \texttt{utils/make\_propagator.py}, is provided with \taurunner{}.
This function instantiates \texttt{PROPOSAL} particle and geometry objects, which are then used to create a propagator instance.
Since \texttt{PROPOSAL} does not support variable density geometries, the \texttt{segment\_body} function is used to segment the \taurunner{} body into a number of constant density layers.
The number of layers is determined by solving for points in the body where fractional change in the density is equal to a constant factor, called \texttt{granularity}.
This argument may be specified by the user, and by default is set to $0.5$.
A single propagator object is created for all $\tau^{\pm}$ and, if needed, for all $\mu^{\pm}$.
Since \taurunner{} assumes $e^{\pm}$ are always absorbed, a propagator will never be made for these.
Whenever a new geometry is used, \texttt{PROPOSAL} creates energy loss tables which are saved in \texttt{resources/proposal\_tables}.
The tables require a few minutes to generate, resulting in an overhead for new configurations, but subsequent simulations with the same geometry will not suffer any slow down.

\subsection{Conventions}
\label{subsec:convent}
\taurunner{} uses a natural unit system in which $\hbar=c=\rm{eV}=1$.
As a consequence of this system, any energy passed to \taurunner{} must be in $\rm{eV}$.
\taurunner{} includes a \texttt{units} package to easily convert common units to the units \taurunner{} expects.
This may be imported from the \texttt{utils} module, and its usage is demonstrated in several examples.
Additionally, since \taurunner{} assumes that propagation occurs in a spherical body, the radius of this body establishes a natural length scale.
Thus all distances are expressed as a fraction of this radius.

\subsection{Output}
\label{subsec:output}
The \texttt{run\_MC} function, which carries out the logic of \taurunner{}, returns a \texttt{numpy.recarray}.
This array may be set to a variable if running \taurunner{} from a script of notebook, or printed or saved if running \taurunner{} from the command line.

In this paragraph, we will describe the fields of this output.
The \texttt{"Eini"} field reports the initial energy of the lepton in $\rm{eV}$.
The \texttt{"Eout"} field reports the energy of the particle when propagation has stopped in $\rm{eV}$.
In the case that the particle was absorbed, this field will always read \texttt{0.0}.
The \texttt{"theta"} field reports the incident angle of the lepton in degrees.
The \texttt{"nCC"} and \texttt{"nNC"} fields report the number of charged and neutral current interactions the particle underwent in its propagation.
The \texttt{"PDG\_Encoding"} field reports the particle type, using the Particle Data Group MC numbering scheme.
The \texttt{"event\_ID"} is a number carried byfield reports which initial lepton the particle comes from.
The \texttt{"final\_position"} field reports the track parameter when the propagation was ended.
This may be used to physical quatities of a particle when it was absorbed, or when a user-defined stopping condition was met



\section{Performance}
\label{sec:performance}
For a given primary spectrum and medium through which to propagate, there are a variety of related factors that determine the runtime of the program, including, but not limited to: (1) the initial energy of the neutrinos, (2) the total column depth of the path, (3) the settings for computing energy losses, and (4) which particles are being tracked.

\begin{figure}
    \centering
    \includegraphics[width=0.48\textwidth]{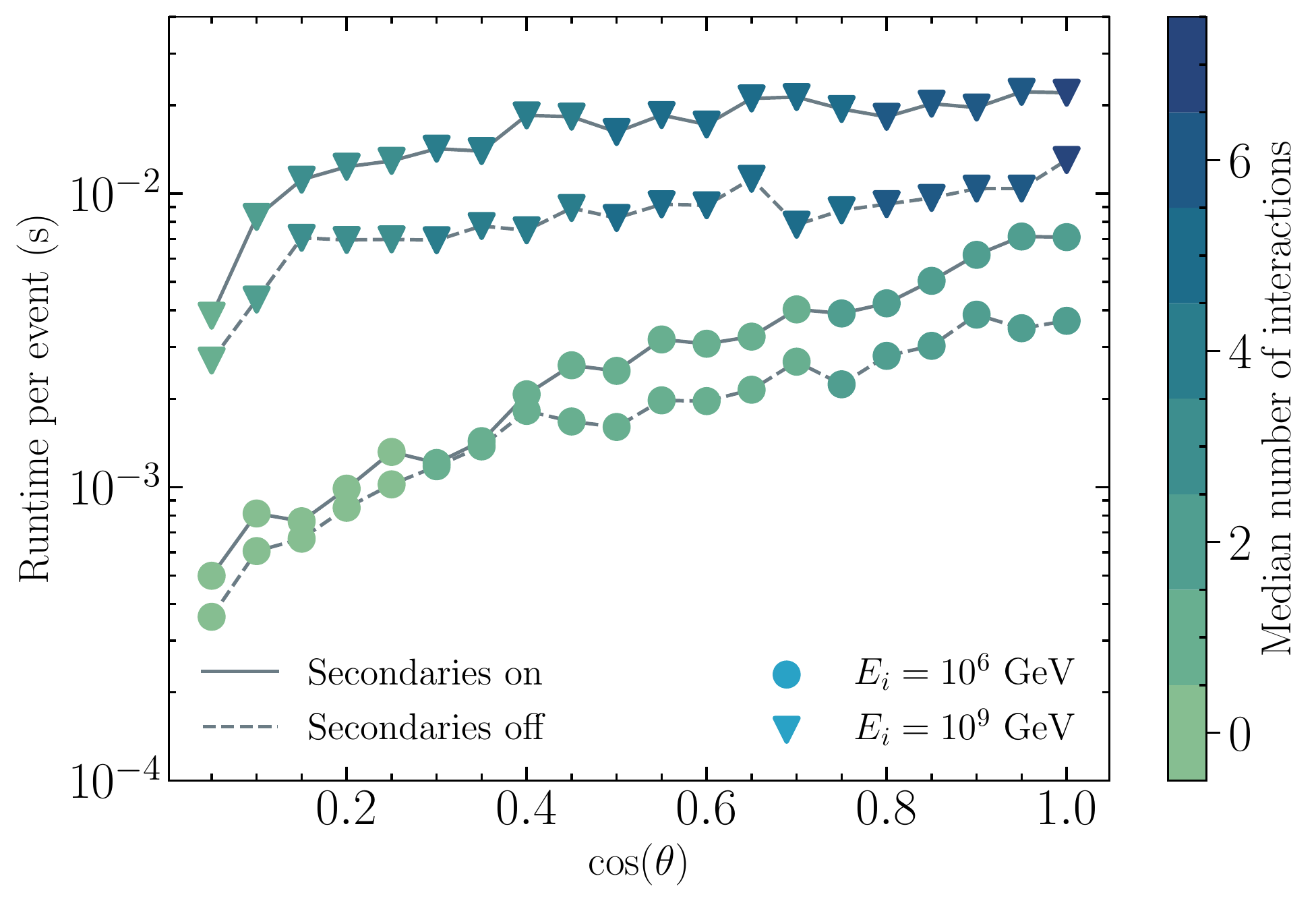}
    \caption{\textbf{\textit{Runtime per $\nu_{\tau}$ event.}} Average runtime per event for various monochromatic fluxes of neutrinos through the Earth, as a function of nadir angle, $\theta$ for incident $\nu_{\tau}$ with energies of 1~PeV (circles) and 1~EeV (triangles). In general, runtime scales with the average number of interactions, which is a function of the energy of the particles and the column depth through which they propagate. The colorbar indicates the median number of NC+CC interactions that the initial beam of $\nu_{\tau}$ undergo. Tracking secondary particles (solid lines) created in $\nu_{\tau}$ CC interactions increases the runtime as the number of particles to propagate increases. Each point represents the average runtime from a simulation including $10^6$ events on a single CPU.}
    \label{fig:performance}
\end{figure}

We show example runtimes for a few different use cases in Fig.~\ref{fig:performance}.
For a fixed \Track{} propagating through Earth, neutrinos with higher initial energy take longer to propagate as they undergo more interactions and as a result experience more stochastic energy losses.
Additionally, those particles that are only being propagated through Earth-skimming trajectories ($\cos(\theta)\approx 0$) can be simulated much quicker than those with large column depths.
This is especially advantageous for proposed Earth-skimming next generation neutrino observatories, \textit{e.g.}~\cite{POEMMA:2020ykm,Neronov:2016zou,GRAND:2018iaj,Aguilar:2019jay,Romero-Wolf:2020pzh}.

By default, all secondary particles that are created as a result of interactions are recorded, meaning that every $\nu_{\tau}$ CC interaction has a chance to increase the number of particles that need to be simulated.
If the user is only interested in outgoing $\nu_{\tau}$ and $\tau$ lepton distributions, this option can be disabled with by setting \texttt{no\_secondaries=True}, which can improve the overall runtime by as much as a factor of two.

Runtime can further be reduced depending on the treatment of energy losses of charged leptons.
By default, energy losses are handled by \texttt{PROPOSAL}~\cite{Koehne:2013gpa}, which treats them stochastically. The user has the choice to ignore energy losses completely, with the setting \texttt{no\_losses=True}, which can improve the runtime by as much as 40\%, although this approximation can only be used in certain scenarios, such as when the initial tau lepton energy is small enough that the interaction length becomes much smaller than the decay length.
This has potential applications for recently proposed indirect searches of ultra-high-energy neutrinos by looking for PeV neutrinos through the Earth~\cite{Safa:2019ege} using large current and next-generation ice or water Cherenkov detectors, such as IceCube-Gen2~\cite{IceCube-Gen2:2020qha}.
Within \texttt{PROPOSAL}, there is also an option to treat energy losses that are below a certain threshold continuously.
We find that setting this parameter to \texttt{vcut=1e-3}, meaning all energy losses that represent less than that fraction of the initial particle energy are treated without stochasticity, achieves an optimal runtime while not neglecting any of the important features that are a result of treating energy losses stochastically.

The first time that a user runs the code, there may be additional overhead while \texttt{PROPOSAL} calculates energy loss distributions for charged leptons.
However, these tables are stored so that future iterations can run more efficiently.
Once the user has run the code at least once and the \texttt{PROPOSAL} energy loss tables are stored, then current runtimes allow users to propagate approximately one million initial EeV $\nu_{\tau}$ through Earth's diameter in approximately eight hours with one CPU.
For an initial energy of one PeV, one million $\nu_{\tau}$ take approximately one hour, depending on the incident angle.
We also found that this runtime varied marginally from machine to machine, and the runtimes in Figure~\ref{fig:performance} and the numbers quoted thus far were all found using a heterogeneous distributed cluster of Linux machines.
The code was also tested on a machine running MacOS with the Apple M1 chip, where the runtimes were found to extremely comparable to those presented above.
For example, $10^4$ $\nu_{\tau}$ with initial energy of one EeV and $\theta=0^{\circ}$ with no secondaries took 0.0127~s per event, on average, and those in the figure above took 0.0124~s per event, on average.

In terms of memory, \taurunner{} can be run on most modern machines, requiring only a few GB of RAM to run.
For example, propagating $10^4$ $\nu_{\tau}$ through the Earth with initial energies of an EeV requires only approximately 1~GB of memory when tracking only $\nu_{\tau}$ and $\tau$, and approximately 3~GB when tracking all particles.
The vast majority of this memory is allocated for calculating energy losses with PROPOSAL, \textit{e.g.} for various trajectories through the Earth and for various initial energies, we found that $\sim 50 - 90\%$ of the memory usage was due to \texttt{PROPOSAL}.
Because most of the memory is due to overhead from the energy losses, there is only a marginal increase in memory usage from propagating many more particles, \textit{e.g.} two sample iterations of the code both took between 2.5~GB and 3.0~GB when propagating $10^4$ or $10^6$ $\nu_{\tau}$ through the Earth with the same initial energies and angles.

\section{Outputs and comparisons}
\label{sec:results}
\begin{figure}[!ht]
    \centering
    \includegraphics[width=0.5\textwidth,trim={0cm 0cm 0cm 0cm}]{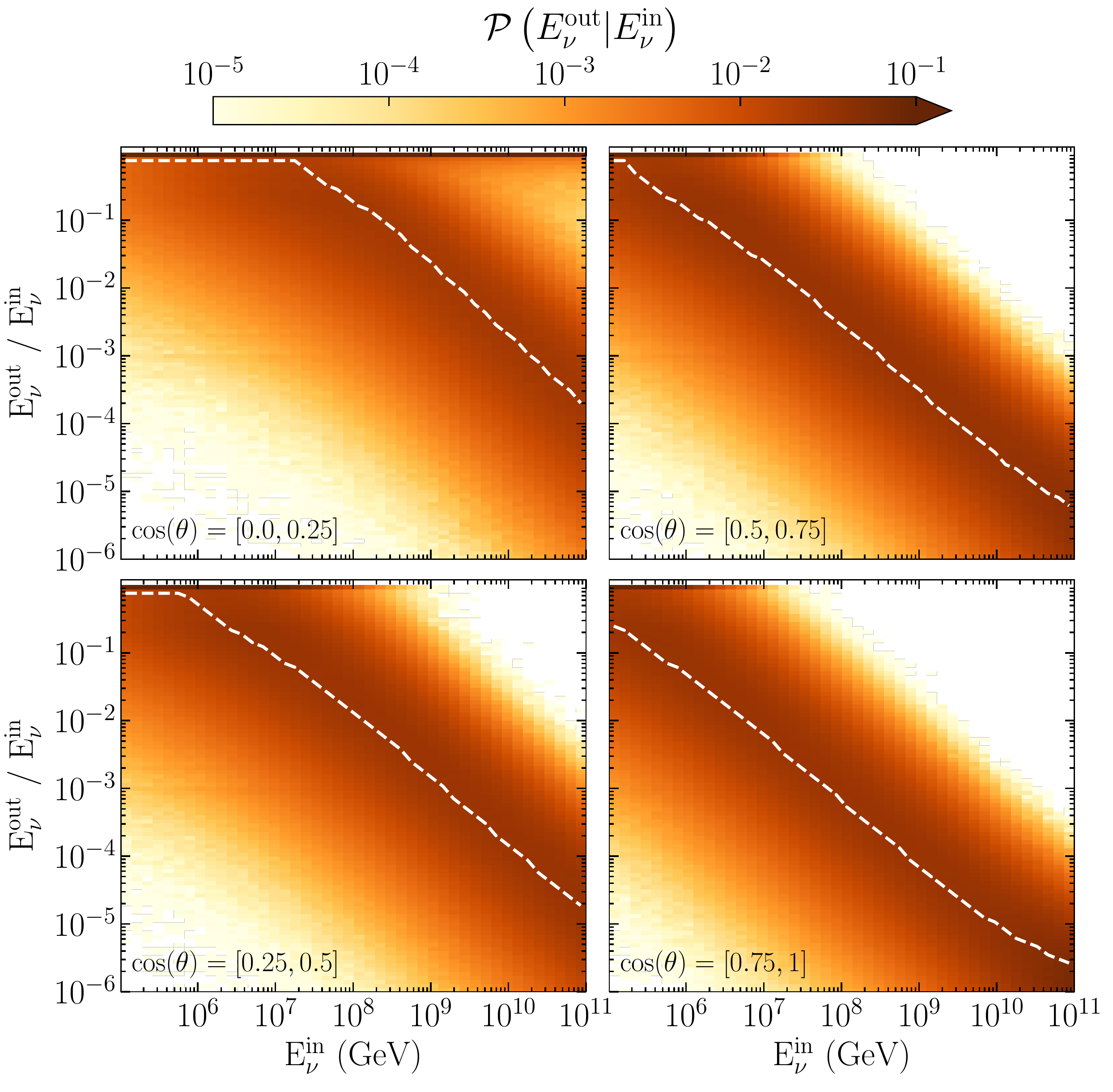}
    \caption{\textbf{\textit{Outgoing $\nu_{\tau}$ distributions for an $E^{-1}$ power-law flux.}} Shown are the outgoing tau neutrino energy fraction as a function of the primary tau-neutrino neutrino flux injected as an E$^{-1}$ power-law from 100 TeV to 10 EeV, shown in slices of equal solid angle in the Northern Sky. Dashed line indicates the median outgoing energy}
    \label{fig:outgoing_grid}
\end{figure}

\begin{figure}[!ht]
    \centering
    \includegraphics[width=0.5\textwidth,trim={0cm 0cm 0cm 0cm}]{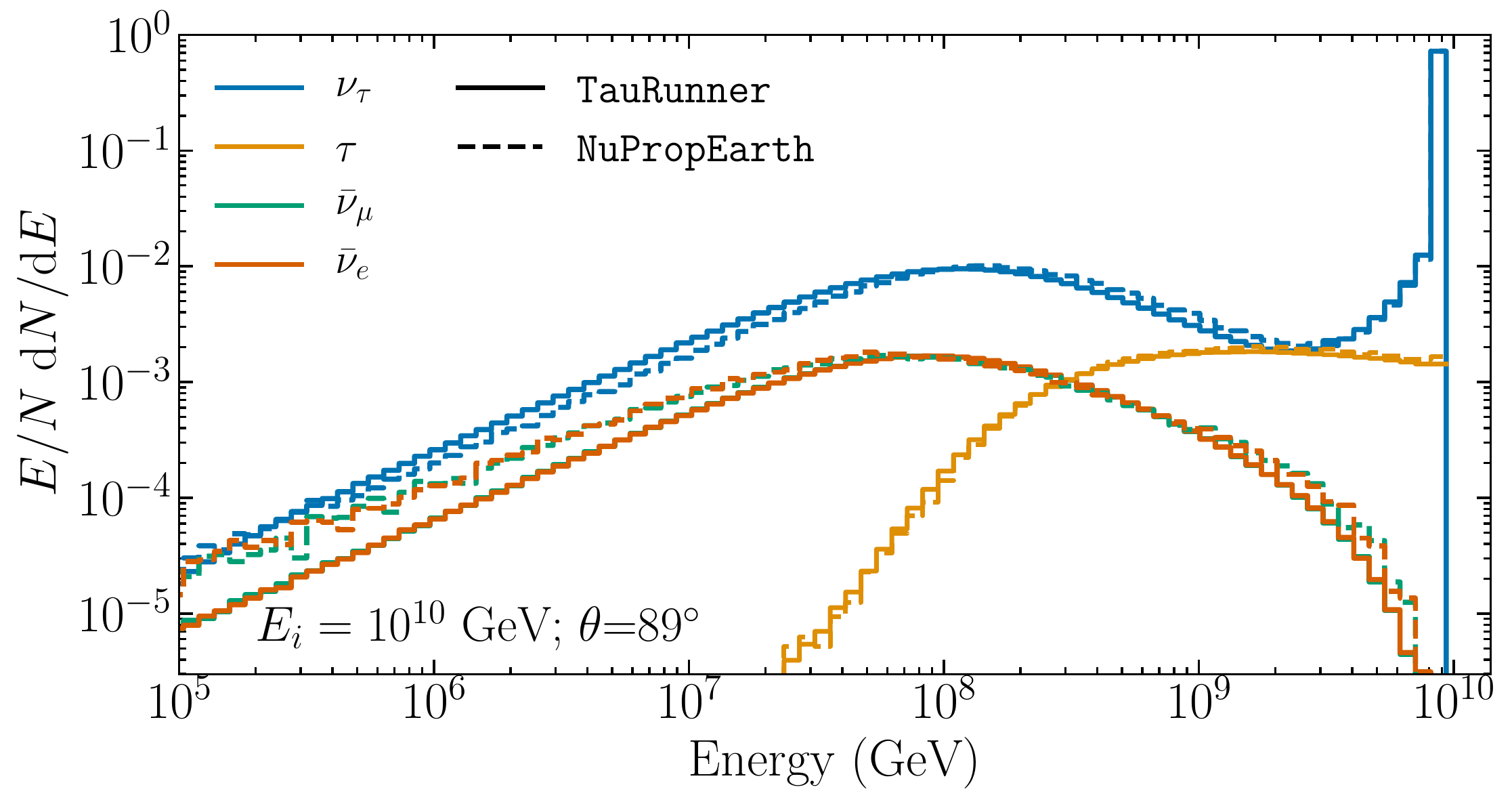}
    \caption{\textbf{\textit{A monochromatic flux of tau neutrinos}}
    Outgoing particle energy distributions for a fixed angle and energy. We include secondary anti-electron and -muon neutrinos, as well as charged taus. \taurunner{} shows good agreement with \texttt{NuPropEarth}. This set assumes Earth as a body with a 4km layer of water.
    }
    \label{fig:outgoing}
\end{figure}

The results of several tau neutrino simulation sets are illustrated in this section. 
Fig.~\ref{fig:outgoing_grid} shows column-normalized distributions of outgoing neutrino energy fraction as a function of initial neutrino energy.
Interestingly, the dashed line showing the median outgoing tau neutrino energy fraction varies with a constant slope, corresponding to the energy at which Earth becomes transparent. 
That energy is roughly $10$ PeV at the horizon (top left), $\mathcal{O}(1)$ PeV in the mantle (top right and bottom left), and $\mathcal{O}(10)$ TeV through the core (bottom right). 
This means that for a large fraction of the Northern Sky, tau neutrinos pile-up and escape at energies where the atmospheric neutrino background is relatively low. 
This idea is also made clear when illustrated for a monochromatic flux. In Fig.~\ref{fig:outgoing}, EeV tau neutrinos are propagated and the outgoing energies are plotted as a function of nadir angle. A similar feature can be seen, where a majority of neutrinos in this simulation escape with energy above $100$ TeV.

\begin{figure}[]
    \centering
    \includegraphics[width=0.45\textwidth,trim={0cm 0cm 0cm 0cm}]{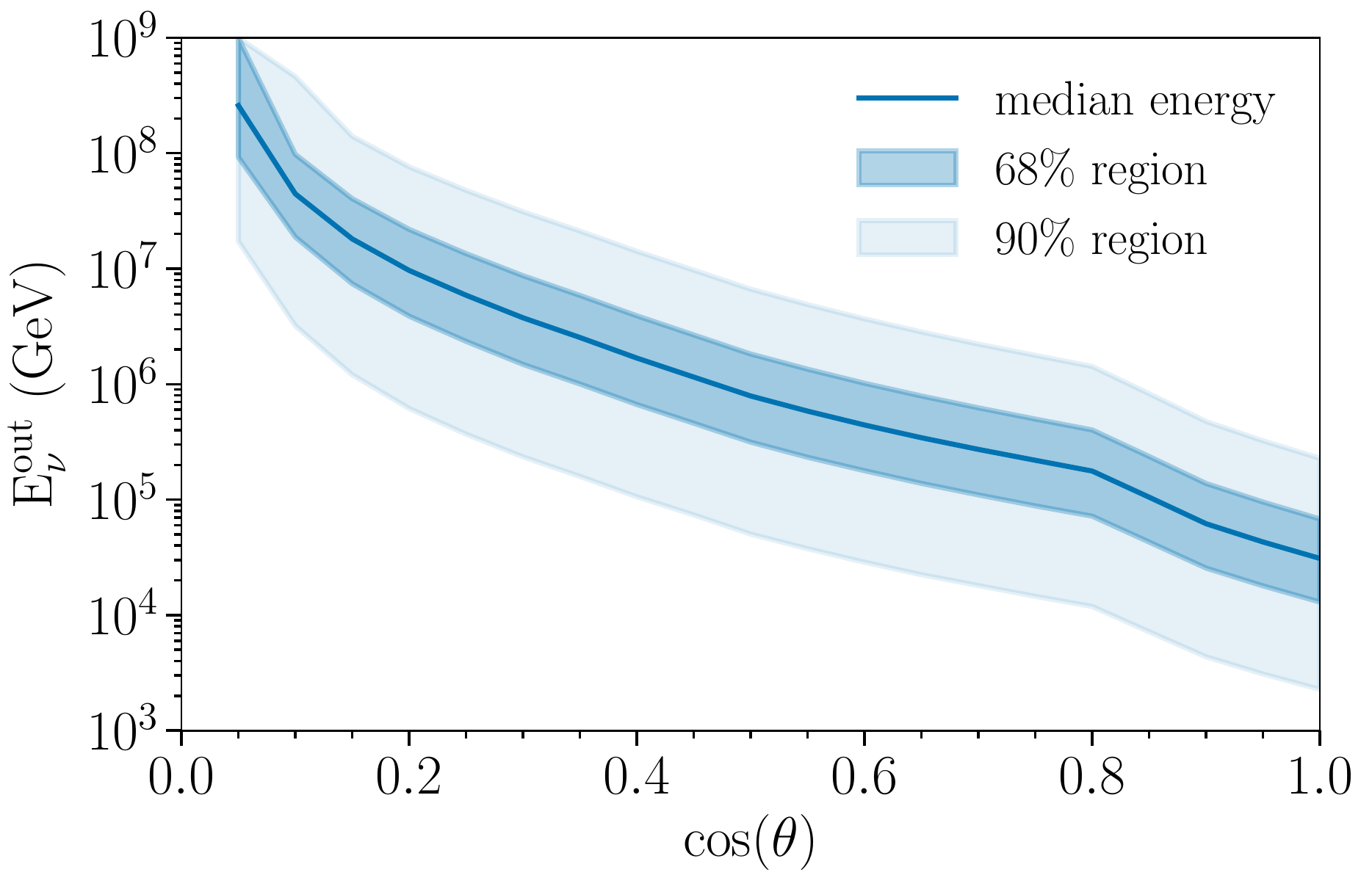}
    \caption{\textbf{\textit{EeV tau neutrinos in Earth}} Median outgoing energies of secondary tau neutrinos shown as a function of nadir angle. Also, $68\%$ and $90\%$ probability contours for outgoing energies are included. The feature at approximately $\cos{\theta}$ of 0.8 is caused by the core.}
    \label{fig:outgoing_theta}
\end{figure}

\taurunner{} has also been compared to several publicly available packages that perform similar tasks.
A summary of the various tested packages and their features is shown in Tab.~\ref{tab:softable}. 
Besides \taurunner{}, only \texttt{NuPropEarth} offers a full solution in the case of tau neutrinos. To illustrate this, we show in Fig.~\ref{fig:outgoing} the output of both packages for an injected monochromatic flux of tau neutrinos at $10^{10}$ GeV and one degree below the horizon. For secondary taus and tau neutrinos, the two packages show excellent agreement. We note that comparisons with \texttt{NuPropEarth} use the trunk version of the code, which has a new treatment for charged particle propagation using \texttt{PROPOSAL} instead of \texttt{TAUSIC}.  Secondary anti-muon and -electron neutrino distributions show slight disagreement in the tails, likely due to different tau polarization treatments. These differences are still being investigated, and will be addressed in an upcoming work.

\begin{figure}[!ht]
    \centering
    \includegraphics[width=0.45\textwidth,trim={0cm 0cm 0cm 0cm}]{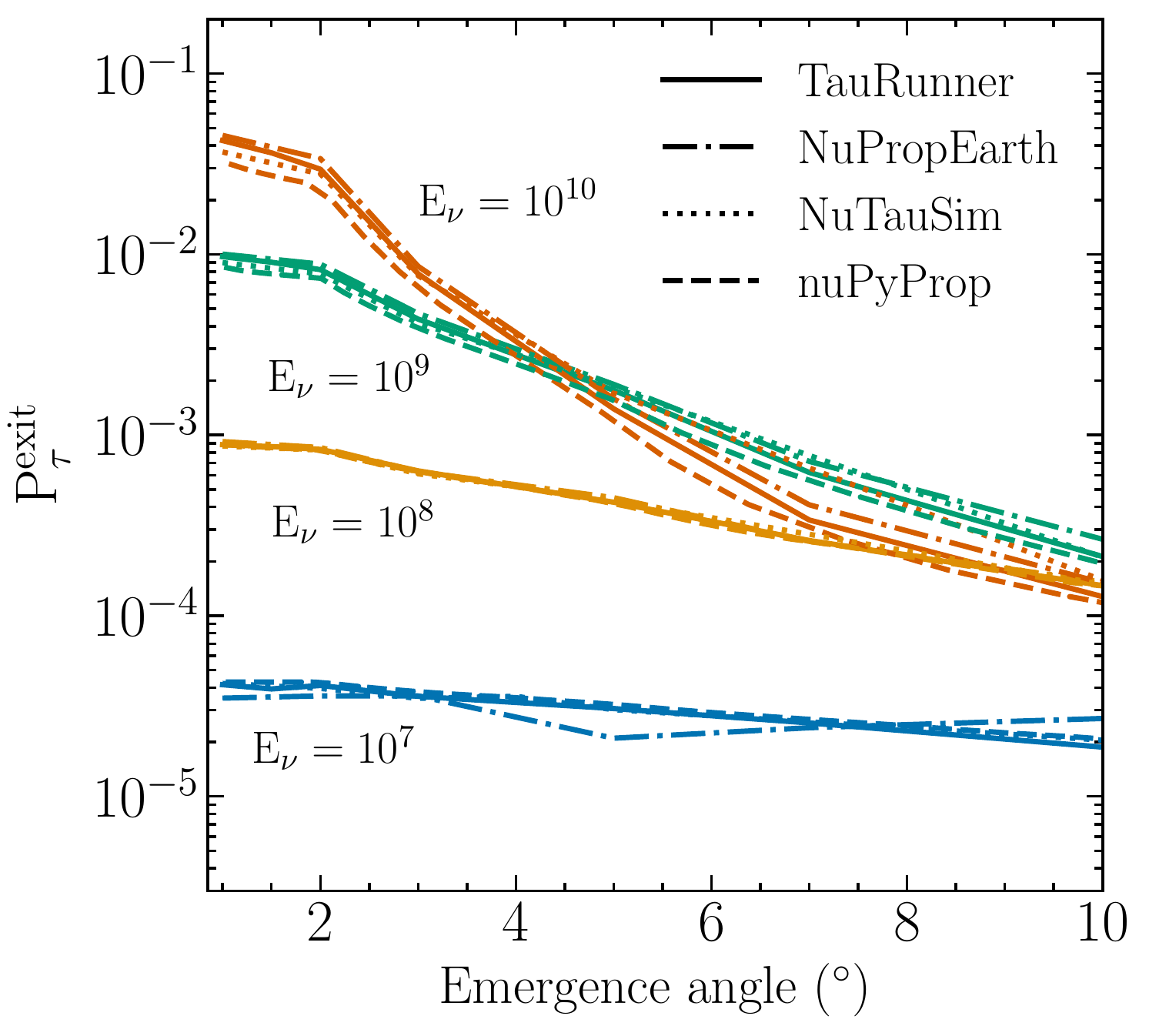}
    \caption{\textbf{\textit{Charged tau lepton exit probability.}} 
    Different colors correspond to four different monochromatic neutrino energies.
    The  emergence angle is measured with respect to horizon.
    The \taurunner{} prediction (solid line) is compared to NuTauSim, NuPropEarth, and nuPyProp, which are shown in different linestyles.}
    \label{fig:pexit_all_softwares}
\end{figure}

Fig.~\ref{fig:pexit_all_softwares} shows a comparison of the charged tau exit probability in Earth as a function of nadir angle. $P_{\rm{exit}}^{\tau}$ is the probability that an incoming neutrino will exit Earth as a charged tau. This quantity is especially relevant for future neutrino observatories hoping to detect Earth-skimming tau neutrinos. In that scenario, exiting taus make up the bulk of the expected signal. \taurunner{} again shows great agreement overall with other packages.

\section{Examples}
\label{sec:examples}
In this section, we show examples which illustrate many of the capabilities of \taurunner{}.
\taurunner{} can be run from the command line or imported as a package.
When a feature can be used via both interfaces, we provide an example for each.

\subsection{Installation}
\label{subsec:install}
\taurunner{} can be installed using \texttt{pip3} by running:

\begin{lstlisting}[language=bash]
pip3 install taurunner
\end{lstlisting}

This will also install any required dependencies, which include \texttt{numpy}~\cite{vanderWalt:2011bqk}, \texttt{scipy}~\cite{Virtanen:2019joe}, and \texttt{PROPOSAL}~\cite{Koehne:2013gpa}.

Furthermore, certain use cases may require access to the source code, which can be downloaded from the \taurunner{} \href{https://github.com/icecube/TauRunner}{GitHub}. After obtaining the source code, one can install the source code with the package manager \texttt{pip3}, while allowing the user to make edits to the source code without having to reinstall the package.

\begin{lstlisting}[language=bash, caption={\textbf{\textit{Installing \taurunner{} using \texttt{pip3} with access to source files}}}]
CLONE_DIR=/path/to/clone/directory
cd $CLONE_DIR
git clone https://github.com/icecube/TauRunner.git && cd TauRunner
pip3 install -e .
\end{lstlisting}

Alternatively, for those that do not use the \texttt{pip3} package manager, one can install all of the dependencies listed in the \texttt{requirements.txt} file included on GitHub, and then clone the repository and add the base directory to the \texttt{PYTHONPATH} variable, as follows:

\begin{lstlisting}[language=bash, caption={\textbf{\textit{Installing \taurunner{} from source.}}}]
CLONE_DIR=/path/to/clone/directory
cd $CLONE_DIR
git clone https://github.com/icecube/TauRunner.git
export PYTHONPATH=$PYTHONPATH:$CLONE_DIR/TauRunner
\end{lstlisting}

\subsection{Monochromatic through Earth}
\label{subsec:monochromatic}
Here we give an example of how to use the most fundamental functionality of \taurunner{}: propagating a monochromatic flux of neutrinos at a fixed energy through a body at a fixed angle.

\begin{lstlisting}[language=Python, caption={\textbf{\textit{Propagating a monochromatic flux from a single angle.}} Example of using an independent script to propagate a monochromatic flux of neutrinos with initial energy $E_{\nu}=10^{10}~{\rm{GeV}}$ from a nadir angle of $89^{\circ}$, \textit{i.e.} one degree below the horizon.}]
import numpy as np

from taurunner.main import run_MC
from taurunner.body.earth import construct_earth
from taurunner.cross_sections import CrossSections
from taurunner.utils import make_propagator, make_initial_e, make_initial_thetas

nevents  = 5000 # number of events to simulate
eini     = 1e19 # initial energy in GeV
theta    = 89.0 # incidence angle (nadir)
pid      = 16 # PDG MC Encoding particle ID (nutau)
xs_model = "CSMS" # neutrino cross section model 

Earth    = construct_earth(layers=[(4., 1.0)]) # Make Earth object with 4km water layer
xs       = CrossSections(xs_model)
energies = make_initial_e(nevents, eini) # Return array of initial energies in eV
thetas   = make_initial_thetas(nevents, theta)

tau_prop = make_propagator(pid, Earth)
rand     = np.random.RandomState(seed=7)

output = run_MC(energies, 
                thetas, 
                Earth, 
                xs,
                tau_prop, 
                rand,
         )
\end{lstlisting}

If you are using the source code installation, you may also achieve this same effect from the command line in the following manner

\begin{lstlisting}[language=bash]
python main.py -n 1000 -e 1e19 -t 89 --xs CSMS -s 7 --save /path/to/outdir/output.npy
\end{lstlisting}

The \texttt{--save} flag tell the program where to save the output.
If this is not specified, the output will be printed as a table.

\subsection{Isotropic Flux through Earth with Power Law Distribution}
\label{subec:powerlaw}
\taurunner{} also allows the user to sample initial neutrino energies from a power law distribution.
For this, the user must provide bounds on the minimum and maximum allowed energies.
Furthermore, the user may sample incidence angles to simulate a isotropic flux.
We demonstrate bot of these features in the following example.

\begin{lstlisting}[language=Python, caption={\textbf{\textit{Multiangle injection with energies drawn from a powerlaw distribution.}} Example of propagating a flux of neutrinos with initial energies sampled from a power law with incidence angles uniformly sampled over a hemisphere.}]
import numpy as np

from taurunner.main import run_MC
from taurunner.body.earth import construct_earth
from taurunner.cross_sections import CrossSections
from taurunner.utils import make_propagator, make_initial_e, make_initial_thetas

nevents        = 5000 # number of events to simulate  
pid            = 16
xs_model       = "CSMS"
no_secondaries = True

Earth = construct_earth(layers=[(4., 1.0)])
xs    = CrossSections(xs_model)
rand  = np.random.RandomState(seed=7)

# Sample power-law with index -2 between 1e6 GeV and 1e12 GeV
pl_exp   = -2 # power law exponent
e_min    = 1e15 # Minimum energy to sample in eV
e_max    = 1e21 # Maximum energy to sample in eV
energies = make_initial_e(nevents,
                          pl_exp, 
                          e_min=e_max, 
                          e_max=e_min, 
                          rand=rand
                         )
                         
# Sample uniform in solid angle over hemisphere
th_min = 0 # Minimum nadir angle to sample from
th_max = 90 # Maximum nadir angle to sample from
thetas = make_initial_thetas(nevents, 
                             (th_min, th_max), 
                             rand=rand
                            )

# tracks   = make_tracks(thetas)
tau_prop = make_propagator(pid, Earth)


output = run_MC(energies, 
                thetas, 
                Earth, 
                xs, 
                tau_prop, 
                rand,
                no_secondaries=no_secondaries
               )
\end{lstlisting}

This may also be accomplished via the command line interface by running:

\begin{lstlisting}[language=bash]
python main.py -n 1000 -e -2 --e_min 1e15 --e_max 1e21 -t range --th_min 0 --th_max 90 -s 7 --xs CSMS

\end{lstlisting}

\subsection{Custom Flux through Earth}
\label{subsec:custom_flux}
The user may also input custom spectra to sample from.
These should be given to \taurunner{} as pickled splines of the flux's cumulative density function.
An example on how to construct these splines in the appropriate format is given in \ref{app:cdf_energy}.
The default \taurunner{} distribution includes splines of different GZK models.
In this example, we show how to sample energies according to the flux predicted in \cite{ahlers:2010fw}.

\begin{lstlisting}[language=python, caption={\textbf{\textit{Propagating a flux drawn from a provided cumulative distribution function (CDF).}} Example of propagating $\nu_{\tau}$ with energies drawn from a user-provided flux. TauRunner provides a few CDFs for the user, or custom CDFs may be built.}]
import numpy as np

import taurunner as tr
from taurunner.main import run_MC
from taurunner.body.earth import construct_earth
from taurunner.cross_sections import CrossSections
from taurunner.utils import make_propagator, make_initial_e, make_initial_thetas

nevents  = 5000
pid      = 16
xs_model = "CSMS"

Earth    = construct_earth(layers=[(4., 1.0)])
xs       = CrossSections(xs_model)

tau_prop = make_propagator(pid, Earth)
rand     = np.random.RandomState(seed=7)

# Sample from pickled CDF
pkl_f    = f'{tr.__path__[0]}/resources/ahlers2010_test.pkl' # Path to pickle file with CDF to sample from
energies = make_initial_e(nevents,
                          pkl_f, 
                          rand=rand
)
                         
# Sample uniform in solid angle over hemisphere   
th_min = 0 # Minimum nadir angle to sample from
th_max = 90 # Maximum nadir angle to sample from  
thetas = make_initial_thetas(nevents, 
                             (th_min, th_max), 
                             rand=rand
)

output = run_MC(energies, 
                thetas, 
                Earth, 
                xs, 
                tau_prop, 
                rand
)
\end{lstlisting}

This may also be accomplished using the command line interface by running:

\begin{lstlisting}[language=bash]
python main.py -n 1000 -e ./resources/ahlers2010_cdf_spline.pkl -t range --th_min 0 --th_max 90 -s 7 --xs CSMS

\end{lstlisting}

\subsection{Radial Trajectory}
\label{subsec:radial}
Besides the chord trajectory, which simulates neutrinos passing through a body from one side to the other, \taurunner{} provides a radial trajectory, which simulates neutrinos originating from the center of a \Body{}.
To use this, one need only modify the call to the \texttt{make\_tracks} function.
Note that the \texttt{theta} argument which was specified previously has no bearing on this, but must be passed due to implementation issues.

\begin{lstlisting}[language=python, caption={\textbf{\textit{Example of propagating $\nu_{\tau}$ along a radial trajectory.}} TauRunner allows for arbitrary particle trajectories. This example shows how to use the \texttt{radial} trajectory, whereas all previous examples have used the \texttt{chord} trajectory.}]
import numpy as np

from taurunner.main import run_MC
from taurunner.body.earth import construct_earth
from taurunner.cross_sections import CrossSections
from taurunner.utils import make_propagator, make_initial_e

nevents  = 5000
eini     = 1e19
pid      = 16
xs_model = "CSMS"

Earth    = construct_earth(layers=[(4., 1.0)]) # Make Earth object with 4km water layer
xs       = CrossSections(xs_model)
energies = make_initial_e(nevents, eini)
thetas   = np.zeros(nevents)

tau_prop = make_propagator(pid, Earth)
rand     = np.random.RandomState(seed=7)

output = run_MC(energies, 
                thetas,
                Earth, 
                xs, 
                tau_prop, 
                rand,
)
\end{lstlisting}

This can also be accomplished from the command line by running:

\begin{lstlisting}[language=bash]
python main.py -n 1000 -e 1e19 -t 89 --xs CSMS -s 7 --track radial
\end{lstlisting}

\subsection{Sun}
\label{subsec:sun}
In addition to the Earth, \taurunner{} allows for propagation in the Sun.
\taurunner{} includes high- and low-metalicity Suns, and a user may provide their own solar model.
We include an example of the form that these solar models should take in Appendix~\ref{app:solar_model}.

\begin{lstlisting}[language=python, caption={\textbf{\textit{Propagating $\nu_{\tau}$ through the Sun.}} Example of how to propagate $\nu_{\tau}$ through a body besides earth.}]
import numpy as np
from taurunner.main import run_MC
from taurunner.body import construct_sun
from taurunner.cross_sections import CrossSections
from taurunner.utils import make_propagator, make_initial_e, make_initial_thetas, units

nevents     = 5000
eini        = 1e13 # the sun is opaque at high energies
theta       = 10.0
pid         = 16
xs_model    = "dipole"
solar_model = "HZ_Sun" # Can also be "LZ_Sun"
                                   
xs       = CrossSections(xs_model)
energies = make_initial_e(nevents, eini)
thetas   = make_initial_thetas(nevents, theta)

sun      = construct_sun(solar_model)
tau_prop = make_propagator(pid, sun, granularity=0.5) 
rand     = np.random.RandomState(seed=7)

output = run_MC(energies, 
                thetas, 
                sun, 
                xs, 
                tau_prop, 
                rand
)
\end{lstlisting}

The same result may be achieved from the command line by running;

\begin{lstlisting}[language=bash]
python main.py -n 1000 -e 2.4e17 -t 45 -s 7 --body HZ_Sun --xs dipole
\end{lstlisting}

\subsection{Constant Slab}
\label{subsec:const_slab}
The user may use the \texttt{radial} track to propagate neutrinos from a `slab' of material of a constant density.
This may be done by making a \texttt{Body} object on the fly in the following manner.

\begin{lstlisting}[language=python, caption={\textbf{\textit{Propagation of $\nu_{\mu}$ through a constant slab.}} Although \taurunner{} only supports spherical bodies, we may use a body of constant density along with a \texttt{radial} trajectory to propagate a particle through a slab of constant density. One may create the slab from the base $Body$ object or use the \texttt{body.slab} object. We do the former here for pedagogical purposes, but we recommend using the latter in practice since it has some computational speed ups.}]
import numpy as np

from taurunner.body import Body
from taurunner.main import run_MC
from taurunner.cross_sections import CrossSections
from taurunner.utils import make_propagator, make_initial_e, make_initial_thetas

nevents  = 5000
eini     = 1e15
theta    = 0
pid      = 14
xs_model = "CSMS"

# Make body with density 3.14 g/cm^3 and radius 1000 km
body     = Body(3.14, 1e3)

xs       = CrossSections(xs_model)
energies = make_initial_e(nevents, eini)
thetas   = make_initial_thetas(nevents, theta)

tau_prop = make_propagator(pid, body)
rand     = np.random.RandomState(seed=7)

output   = run_MC(energies, 
                  thetas, 
                  body, 
                  xs, 
                  tau_prop, 
                  rand,
                  flavor=pid
)
\end{lstlisting}

\subsection{Layered Slab}
\label{subsec:layered_slab}
The constant density slab may be generalized to a slab of multiple layers.
As mentioned in Sec.~\ref{subsec:track}, the densities in each layer may be positive scalars, unary functions which return positive scalars, or a potentially mixed list of such objects.
In this example, we show how to accomplish this latter option.

\begin{lstlisting}[language=python, caption={\textbf{\textit{Propagation of $\nu_{\tau}$ through a layered slab.}} We may employ the same strategy of using a \texttt{radial} trajectory to replicate propagation through a slab to propagate through a slab with varying properties.}]
import numpy as np

from taurunner.body import Body
from taurunner.main import run_MC
from taurunner.cross_sections import CrossSections
from taurunner.utils import make_propagator, make_initial_e, make_initial_thetas

nevents  = 1000
eini     = 1e15
theta    = 0
pid      = 16
xs_model = "CSMS"

# Make layered body with radius 1,000 km
def density_f(x):
    return x**-2/4
densities  = [4, density_f, 1, 0.4]
boundaries = [0.25, 0.3, 0.5, 1] # Right hand boundaries of the layers last boundary should always be 1
body = Body([(d, b) for d, b in zip(densities, boundaries)], 1e3)

xs       = CrossSections(xs_model)
energies = make_initial_e(nevents, eini)
thetas   = make_initial_thetas(nevents, theta)

tau_prop = make_propagator(pid, body)
rand     = np.random.RandomState(seed=7)

output = run_MC(energies, 
                thetas, 
                body, 
                xs, 
                tau_prop, 
                rand
)
\end{lstlisting}

\section{Conclusion}
\label{sec:conclusion}
In this article, we have introduced a new package to propagate high-energy neutrinos in a variety of scenarios.
Our implementation includes the dominant neutrino-propagation effects and is valid in the energy range of current and proposed neutrino telescopes.
Additionally, in our performance section, we have compared our package with other state-of-the-art solutions to this problem and find them in good agreement where they overlap.
Finally, the \taurunner{} package is designed to be extendable by the user, by either providing improved or altered physics inputs or constructing new geometries, giving the user the ability to extend the package functionality beyond the examples provided in this article.
The authors hope that this work will encourage further development of publicly available physics software.

\section*{Acknowledgements}
We acknowledge useful discussions with Joseph Farchione, Alfonso Garcia-Soto, Austin Lee Cummings, Andres Romero-Wolf, and Kareem Ramadan Hassan Aly Muhammad Farrag.
We additionally thank Hallsie Reno, Sameer Patel, and Diksha Garg for insightful discussions on tau physics.
We further thank Christopher Weaver for providing updated cross section tables and engaging discussions on non-trivial interpolation problems.
We would also like to thank Gwenha{\"e}l de Wasseige for providing the solar models used in this work.
Finally, we would like to give special acknowledgment to Francis Halzen for his support and discovery of tau regeneration, which was pivotal to this work.
IS, JL, AP, and JV are supported by NSF under grants PLR-1600823 and PHY-1607644 and by the University of Wisconsin Research Council with funds granted by the Wisconsin Alumni Research Foundation.
OV acknowledges support by the Harvard College Research Program in the fall of 2020.
CAA is supported by the Faculty of Arts and Sciences of Harvard University, and the Alfred P. Sloan Foundation.

\bibliography{taurunner.bib}

\pagebreak
\appendix

\section{Constructing CDFs from which to Sample}
\label{app:cdf_energy}
\taurunner{} offers the user the capability to provide custom spectra from which to sample initial energies.
In this appendix, we describe the form in which \taurunner{} expects these spectra, and provide an example of constructing one.
These should be \texttt{scipy.interpolate.UnivariateSpline} objects whose $x$-axis is the value of the cumulative density function of the spectra to sample and whose $y$-axis is the true neutrino energy in $\rm{eV}$.
We now provide an example of constructing these splines. The \texttt{.csv} file we use for this contains one column of energies in GeV and a corresponding column of the squared energies times the number density of the flux in units of GeV. 
It may be found at \texttt{resources/ahlers2010.csv}.

\begin{lstlisting}[language=python, caption={\textit{\textbf{Constructing custom flux files in the format required by \taurunner{}~.}}}]
import numpy as np
from scipy.integrate import quad
from scipy.interpolate import UnivariateSpline
import pickle

import taurunner as tr
from taurunner.utils import units

# csv of a benchmark GZK flux
infile = f'{tr.__path__[0]}/resources/ahlers2010.csv'
tab_data = np.genfromtxt(infile, delimiter = ',')
gzk_e    = tab_data[0]*units.GeV # Convert energies to eV
gzk_dnde = tab_data[0]*units.GeV

gzk_mine = gzk_en[0]
gzk_maxe = gzk_en[-1]

# Splining in logspace recommended 
gzk_spline = UnivariateSpline(np.log(gzk_en), np.log(gzk_flux/gzk_en**2), k = 4, s=1e-2)

integrand = lambda E: np.exp(gzk_spline(np.log(E)))

# integrating in logspace also recommended
norm, _ = quad(lambda x: np.exp(x)*integrand(np.exp(x)), np.log(gzk_min), np.log(gzk_max))

pdf = lambda E: integrand(E) / norm

# Make and spline CDF
cdf_energies = np.logspace(np.log10(gzk_min), np.log10(gzk_max*1.1), 500) # Maybe more knots than necessary but more support is better
cdf = np.array([integrate.quad(lambda x: np.exp(x)*probability(np.exp(x)), np.log(gzk_min), np.log(y))[0] for y in cdf_energies])
# Make sure this in invertible
mask = np.where(np.logical_and(cdf>0, cdf<=1))[0]
cdf = cdf[mask]
cdf_energies = cdf_energies[mask]
cdf_spl = UnivariateSpline(cdf, cdf_energies)

# Save the spline as a pickle file
out_f = f'{tr.__path__[0]}/resources/ahlers2010.pkl'
with open(out_f, 'wb') as pkl_f:
    pkl.dump(cdf_spl, pkl_f)
\end{lstlisting}

Saving the file in resources is not necessary.
The user may now sample from this distribution by passing the path to the file as the energy argument in the command line or as the first argument of the \texttt{make\_initial\_e} function seen in the examples.
A more detailed example of constructing these splines in a \texttt{Jupyter Notebook} along with some sanity checks may be found on our GitHub in the \texttt{examples} folder.

\section{Cross Section Splines}
\label{app:xs}
In this section we give an example of saving cross section splines in the form required by \taurunner{} so that the user may pass their own cross section model if they so choose.
The differential splines should be a \texttt{scipy.interpolate.RectBivariateSpline} object and the total cross section splines should be a \texttt{scipy.interpolate.UnivariateSpline} object.
We will now work out an example, assuming that we have two \texttt{.csv} files, one each for total and differential cross sections.
In the former case, we will assume that it has two columns, the first containing neutrino energies and the second the corresponding total cross section.
In the latter case, we will assume that we have three columns, the first containing an incoming neutrino energy, the second containing convenience variable described in Sec.~\ref{subsec:xs}, and the third containing the corresponding differential cross section.
All energy units will be assumed to be $\rm{GeV}$ and all area units $\rm{cm}^{2}$.
In the case of the differential cross section, the values of the convenience variable must be the same for each incoming neutrino energy.
As a reminder, \taurunner{} assumes that the cross section is the same for all neutrino flavors and thus one need only make only one set of cross section splines.

\begin{lstlisting}[language=python, caption={\textbf{\textit{Example of constructing differential cross section splines for \taurunner{}.}}}]
import numpy as np
from scipy.interpolate import UnivariateSpline
import pickle

import taurunner as tr
from taurunner.utils import units

model_name  = "my_model"
interaction = "CC" # Charged current
nucleon     = "p" # proton
nutype      = "nubar" # antineutrino

# csv containing the anti-neutrino proton CC xs
tot_xs_path = f"/path/to/{nutype}_{nucleon}_{interaction}_xs.csv"
e = np.genfromtxt(tot_xs_path, delimiter=",")[0]
xs = np.genfromtxt(tot_xs_path, delimiter=",")[1]

# Convert to natural units
e = e*units.GeV
xs = xs*units.cm**2

# Spline in logspace
xs_spl = UnivariateSpline(np.log(e), np.log(xs))

# Save the spline as a pickle file
# Splines must follow this naming convention and be in this directory
out_f = f"{tr.__path__[0]}/resources/cross_section_tables/{model_name}_{nutype}_{nucleon}_sigma_{interaction}.pkl"
with open(out_f, "wb") as pkl_f:
    pkl.dump(xs_spl, pkl_f)

\end{lstlisting}

This process would then be repeated for all combinations of interaction type $\in$ [\texttt{"CC"}, \texttt{"NC"}], neutrino type $\in$ [\texttt{"nu"}, \texttt{"nubar"}], and nucleon $\in$ [\texttt{"p"}, \texttt{"n"}] for a total of 8 splines.
Now we show a similar example for constructing differential cross section splines.
\taurunner{} splines have support down to $1~\rm{GeV}$, and this number is used internally.
While it is not strictly necessary to have support down to this energy, it is possible that \taurunner{} may evaluate the splines in this regime, and thus understanding the behavior of splines in this regime is recommended.

\begin{lstlisting}differential
import numpy as np
from scipy.interpolate import RectBivariateSpline
import pickle

import taurunner as tr
from taurunner.utils import units

model_name  = "my_model"
interaction = "NC" # Neutral current
nucleon     = "n" # neutron
nutype      = "nu" # neutrino

# csv containing the neutrino neutron NC dsigma/de
tot_xs_path = f"/path/to/{nutype}_{nucleon}_{interaction}_dsde.csv"

e_in = np.genfromtxt(tot_xs_path, delimiter=",")[0]
z    = np.genfromtxt(tot_xs_path, delimiter=",")[1]
dsde = np.genfromtxt(tot_xs_path, delimiter=",")[2]

# Convert to natural units
e_in  = e_in*units.GeV
dsde  = dsde*units.cm**2/units.GeV
dsdx  = dsde*e_in

# Spline in logspace
xs_spl = RectBivariateSpline(np.log(np.unique(e_in)), z, np.log(dsdx))

# Save the spline as a pickle file
# Splines must follow this naming convention and be in this directory
out_f = f"{tr.__path__[0]}/resources/cross_section_tables/{model_name}_{nutype}_{nucleon}_dsde_{interaction}.pkl"
with open(out_f, "wb") as pkl_f:
    pkl.dump(xs_spl, pkl_f)

\end{lstlisting}

As in the case of the total cross section, this process must be repeated for all combinations of interaction type $\in$ [\texttt{"CC"}, \texttt{"NC"}], neutrino type $\in$ [\texttt{"nu"}, \texttt{"nubar"}], and nucleon $\in$ [\texttt{"p"}, \texttt{"n"}] for a total of 8 splines.
This new model may then be used by passing \texttt{"my\_model"} when initializing the \XS{} object.

\section{Solar Model Format}
\label{app:solar_model}
\taurunner{} expects solar models to have at minmum three columns, one containing the radius in units of the solar radius, one containing the corresponding mass density in ${\rm{g}}/\rm{cm}^{3}$, and the last containing the corresponding electron density in $N_{A}^{-1}\rm{cm}^{-3}$.
These values should not be comma separated and lines beginning with \texttt{\#} will be ignored as comments.
Any additional columns will be ignored by \taurunner{}, allowing the user to add additional columns if it is useful, for \textit{e.g.} a column containing the proton fraction to pass to the body.

\end{document}